\documentclass[aps,twocolumn,showpacs,preprintnumbers,floatfix, prb, superscriptaddress]{revtex4-2}

\usepackage{graphicx}  
\usepackage{subfigure}
\usepackage{multirow}

\usepackage{amssymb}

\usepackage{fancyhdr}
\usepackage{longtable}
\usepackage[OT1]{fontenc}
\usepackage{dcolumn}   

\usepackage{bm}        
\usepackage{amsfonts}  
\usepackage{amsmath}   
\usepackage{amssymb}   
\usepackage{xcolor}
\usepackage{soul}
\usepackage{nicefrac}

\newcommand{\dd}{\mathrm{d}}

\newcommand{\pwisein}{\left\{ \begin{array}{ll}}
\newcommand{\pwiseout}{\end{array}\right.}

\usepackage{xcolor}

\begin{document}

\title{Spatial correlations in SIS processes on  \textcolor{black}{random regular} graphs}

\author{Alexander \surname{Leibenzon}}
\affiliation{Racah Institute of Physics, Hebrew University of Jerusalem, Jerusalem 91904, Israel}
\author{Samuel W.S. \surname{Johnson}}
\author{Ruth E. \surname{Baker}}
\affiliation{Wolfson Centre for Mathematical Biology, Mathematical Institute, University of Oxford}
\author{Michael \surname{Assaf}}
\affiliation{Racah Institute of Physics, Hebrew University of Jerusalem, Jerusalem 91904, Israel}

\begin{abstract}  
In network-based SIS models of infectious disease transmission, infection can only occur between directly connected individuals. This constraint naturally gives rise to spatial correlations between the states of neighboring nodes, as the infection status of connected individuals becomes interdependent. 
\textcolor{black}{Although mean-field approximations and the standard pairwise model are commonly used to simplify disease forecasting on networks, they inadequately capture spatial correlations; mean-field frameworks assume that populations are well-mixed, while the pairwise model neglects correlations beyond nearest-neighbor connections, which leads to inaccurate predictions of infection numbers over time.}
As such, the development of \textcolor{black}{approximations} that account for \textcolor{black}{higher order} spatially correlated infections is of great interest, as they offer a compromise between accurate disease forecasting and analytic tractability. 
Here, we use existing corrections to mean-field theory on the regular lattice to construct a more general framework for equivalent corrections on random regular graph topologies. 
We derive and simulate a \textcolor{black}{hierarchical} system of ordinary differential equations for the time evolution of the spatial correlation function at various geodesic distances on random networks. Solving these equations allows us to predict the time-dependent global infection density, which agrees well with numerical simulations. 
Our results substantially improve on existing corrections to mean-field theory for infectious individuals in SIS processes and provide an in-depth characterization of how structural randomness in networks affects the dynamical trajectories of infectious diseases on networks. 

\end{abstract}

\maketitle  

\section{Introduction} 
\begin{figure*}
    \centering
    \includegraphics[width=.87\linewidth]{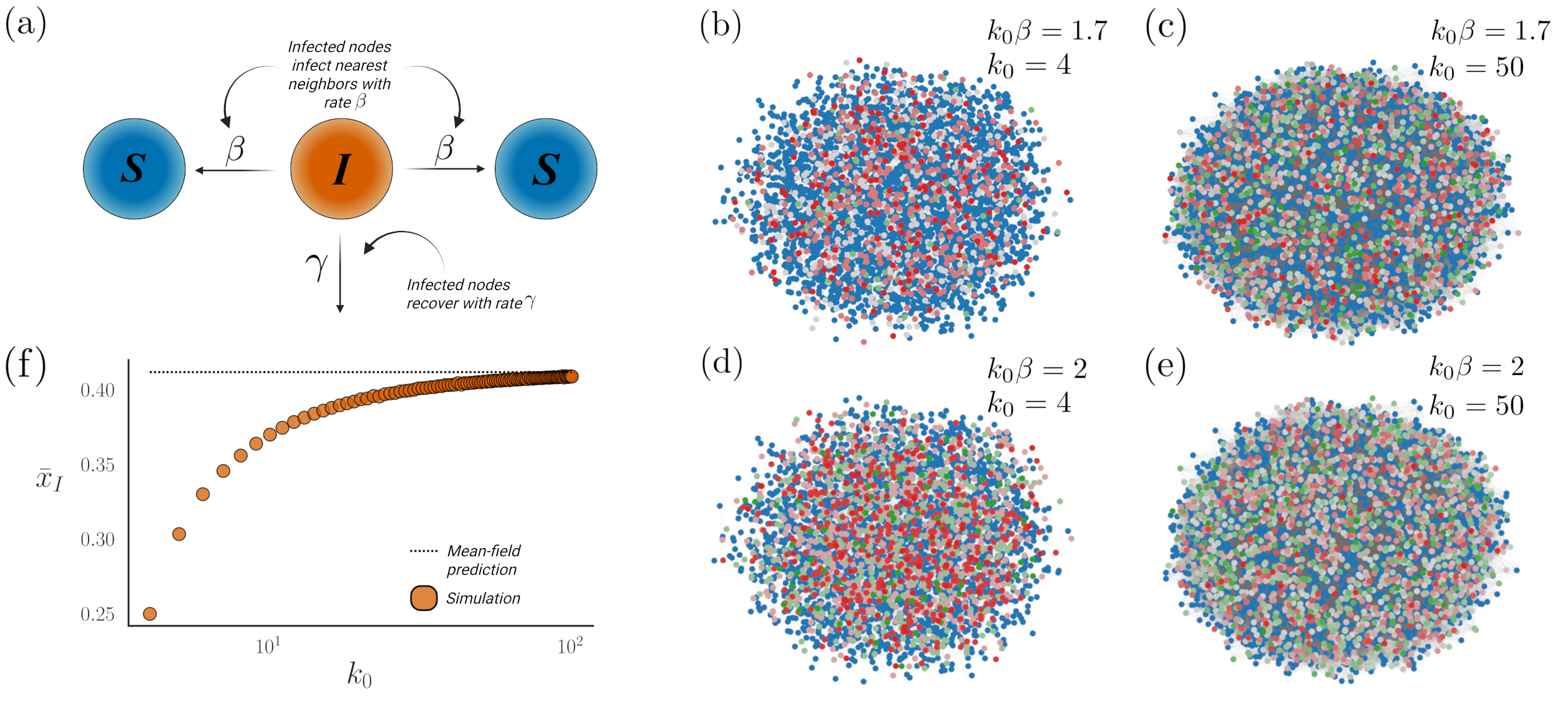}
    \vspace{-6mm}\caption{(a) Schematic of the SIS model on networks. Infected nodes recover with rate $\gamma$ (conventionally, time is rescaled as $t \to \gamma t$, implying $\gamma = 1$) and infect their nearest neighbors, each at rate $\beta$. (b)-(e) Snapshots of SIS model simulations on RRGs \textcolor{black}{with $N=4000$ nodes} in the steady state. \textcolor{black}{The left column [(b), (d)] corresponds to a node degree $k_0=4$ and the right column [(c), (e)] to $k_0=50$; in addition, in (b) and (c)  $\beta k_0=1.7$, and in (d) and (e) $\beta k_0=2$. Susceptible nodes are blue, whereas infectious nodes are colored by their local deviation from the mean-field expectation: red, gray, and green denote positive, zero, and negative spatial correlations, respectively. Stronger positive correlation is observed at lower $k_0$, while at higher $k_0$ a symmetric, uncorrelated mix of red and green fluctuations is apparent. Increasing the infection rate, $\beta$, also decreases these local correlations but also increases the overall infection density in the steady state.} 
    (f) Steady state infection density versus $k_0$, obtained from numerical simulations on RRGs with $N=10^4$ nodes, $\beta k_0 = 1.7$, and $\gamma = 1$. The dashed line is the expected mean-field value of $1-1/(\beta k_0) \simeq 0.41$. \textcolor{black}{Each data point represents an average over 100 realizations.} In all cases, $100$ infectious individuals are initially placed randomly on the network, \textcolor{black}{ with all other nodes being susceptibles.} }
    \label{fig:schematic}
\end{figure*}

The Susceptible-Infected-Susceptible (SIS) model, {\color{black} rooted in the compartmental framework of Kermack and McKendrick~\cite{kermack1932contributions}}, is a foundational framework for describing the spread of infectious diseases in which individuals do not acquire lasting immunity following infection (e.g. the common cold or many sexually transmitted diseases~\cite{da2008global, turner1997epidemiology, eames_modeling_2002, eames2004monogamous}).
In the model, members of a population transition repeatedly between two states: susceptible (\textit{S}) and infected (\textit{I}). \textcolor{black}{The dynamics of the SIS model are governed by two processes: the infection per connection between infected and susceptible individuals, occurring at a rate $\beta$, and the recovery of infectious individuals at a rate $\gamma$ (see Fig.~\ref{fig:schematic}(a)).} Many generalized SIS models assume that susceptible and infected populations are well-mixed spatially, such that \textcolor{black}{the rates of contact between them} are assumed to be uniform across a population. 
Such models are often referred to as \textit{mean-field} approximations and, while they are highly amenable to mathematical analysis~\cite{seno2020sis, real2007spatial, eames_modeling_2002}, they are generally inappropriate for real-world scenarios, where infected individuals are spatially ordered or clustered. 
In such contexts, the dynamical trajectory of a disease is highly sensitive to the spatial distribution of infected individuals, with the infection spreading more rapidly when infected individuals are distributed more evenly in space~\cite{keeling_effects_1999}.

One solution to mitigate the inaccuracies of mean-field models is to represent the SIS process on a network, in which nodes represent individuals in a population and edges represent connections between individuals (i.e., potential routes for the flow of infection). \textcolor{black}{This explicit connectivity necessarily leads to the emergence of spatial correlations between infectious individuals, as transmission can only occur via direct links to currently infected neighbors (see example in Figs.~\ref{fig:schematic}(b)-(e)).} This local transmission mechanism leads to epidemic trajectories that differ markedly from those predicted by mean-field models. This is especially evident on networks with lower average node degree, \textcolor{black}{in which these deviations are more pronounced (Fig.~\ref{fig:schematic}(f)), or on regular lattices with a high degree of spatial order. A significant analytical advance in capturing these dynamics} was the development of pairwise approximation methods. \textcolor{black}{By tracking the densities of connected node pairs of various types (e.g., S-I pairs), these models successfully account for these emergent local correlations. Consequently, pairwise frameworks are more accurate than standard mean-field models, particularly in the aforementioned low-degree regimes and on heterogeneous topologies~\cite{keeling_effects_1999, keeling_correlation_1997, eames_modeling_2002}.}


 
Despite these improvements, existing ODE-based approaches rely on closure approximations that assume spatial correlations between infectious individuals decay rapidly with shortest path distance~\cite{keeling_effects_1999, baker2010correcting, markham2013simplified}. \textcolor{black}{The standard pairwise model (SPM)~\cite{keeling_effects_1999, kiss2017mathematics} is inherently limited by this; it restricts its pair-level description exclusively to neighboring nodes, and neglects spatial correlations at larger distances between non-neighboring infected individuals. Thus, no analytical framework exists that systematically accounts for multi-scale spatial correlations on generic networks, and in particular, on random regular networks (RRGs), making analysis challenging and leading to a preference for simulation-based exploration.}

To address this challenge, this work extends \textcolor{black}{the SPM} to derive a general framework for the accurate forecasting of disease spread via SIS processes on RRG topologies. Motivated by numerical simulations that highlight the strong dependence of spatial correlations on underlying network structure, particularly at lower node degrees, we first analyze spatial correlations in infection dynamics on a series of regular networks with increasingly randomized connectivity. 
Focusing specifically on the limiting case of the  RRG, we then generalize existing lattice-based theory, deriving a system of ODEs to model the temporal evolution of spatial correlations at varying shortest path lengths. 
\textcolor{black}{Unlike the SPM,} our multi-shell pairwise model (MPM) resolves the full hierarchy of spatial correlations at every network distance by solving \textcolor{black}{a system of coupled ODEs describing the dynamics within each shell (i.e., the set of nodes at a given shortest-path distance)} around an infected node. This enables us to capture how infection clustering decays with distance and how finite network connectivity and network structure shape epidemic trajectories, which are missed by standard closures.
Comparison with numerical simulations demonstrates that our derived methodology yields substantially improved accuracy over mean-field and SPM predictions.  Furthermore, by systematically refining mean-field theory with spatial correction terms that account for local clustering, we derive closed-form asymptotic expansions for both the endemic infection density and the pairwise spatial correlations in the long-time limit. These steady-state approximations not only quantify the impact of network structure on endemic persistence, but also provide expressions that show how finite connectivity suppresses infection prevalence, thereby reducing the accuracy of classical mean-field predictions. Thus, our results offer a robust analytical tool for understanding and forecasting the dynamics of infectious diseases driven by SIS processes on RRGs, revealing how structural randomness affects the spatial organization of infections and thereby shapes epidemic trajectories.

\section{Corrections to SIS mean-field theory on regular lattices}
\label{existingCorrections}

The mean-field mathematical representation of SIS dynamics executed on the RRG makes the simplifying assumption that susceptible and infected populations are well-mixed spatially, permitting the construction of two simple ODEs governing the dynamics of the system
\begin{align}
    \frac{\dd x_{S}}{\dd t} = - \beta k_0 x_{I}x_{S} + \gamma x_{I},\;\;\;
     \frac{\dd x_{I}}{\dd t} = \beta k_0 x_{I}x_{S} - \gamma x_{I}, \label{originalODE}
\end{align}
where $\beta>0$ is the rate of infection of one susceptible individual from one infectious individual, $\gamma>0$ is the recovery rate of each infectious individual, $k_0$ is the fixed degree of each node, and $0\leq x_{I},x_{S} \leq1$  are the population fractions in the $I$ and $S$ states, respectively. 

The conservation law $x_{S} + x_{I} = 1$ means that every individual is either susceptible or infected at any point in time, and time is often rescaled as $t\to \gamma t$ by setting $\gamma = 1$. Hence, Eq.~\eqref{originalODE} can be reduced to 
\begin{equation}
    \frac{\dd x_{I}}{\dd t} = \beta k_0 x_{I}(1 - x_{I}) - x_{I}. 
    \label{eq:SISLogistic}
\end{equation}
In Eq.~\eqref{eq:SISLogistic}, the first term represents the infection at a rate $\beta$ of  susceptible nodes by an average of $k_0x_I$ infected neighbors,  while the second term represents the recovery of infectious individuals at a rate of $1$. 

\begin{figure}[t!]
    \centering
    \includegraphics[width=0.84\linewidth]{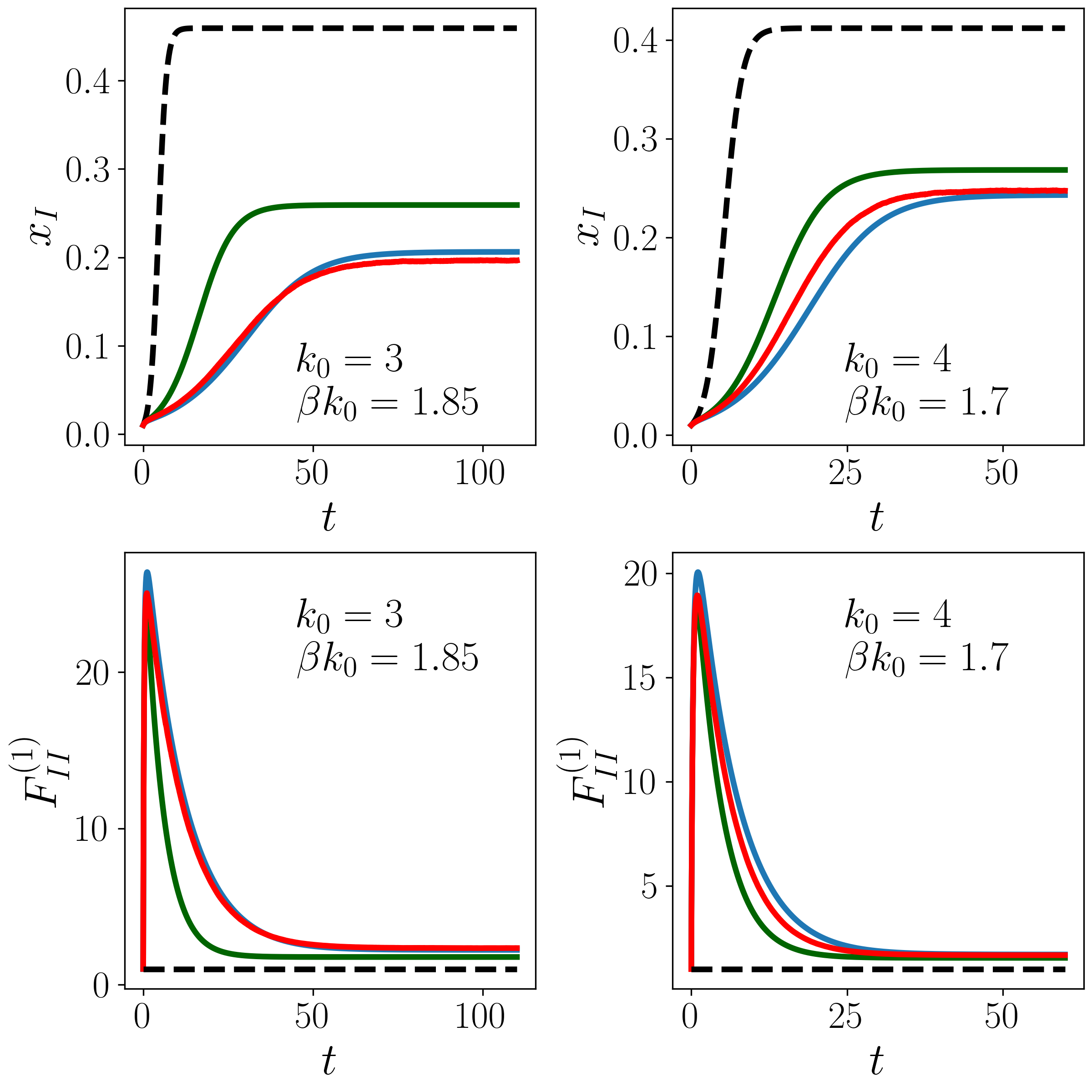}
    \vspace{-.5cm}\caption{Average infection density, $x_{I}$ (top row), and pairwise correlation function, $F_{II}^{(1)}$ (bottom row). Here, SIS model simulations on the RRG (red) are compared with numerical solutions of the \textcolor{black}{MPM (blue), SPM (dark-green),} and mean-field theory (dashed black). Simulations were executed on networks with $N=10^4$ nodes and $\gamma=1$, \textcolor{black}{and in all simulations $100$ infectious individuals are initially placed randomly on the network, with all other nodes occupied by susceptible individuals, and all results are averaged over 100 simulations.} {\color{black} The left column corresponds to  $k_0=3$ with $\beta k_0=1.85$, while the right column corresponds to $k_0=4$ with $\beta k_0=1.7$.}}    \label{fig:RRG_Vailidity}
\end{figure}

Under this mean-field assumption, $x_{I}(1 - x_{I})$ gives the probability that an individual at a randomly selected node is susceptible and a specific neighbor is infected, which follows from the assumption that all nodes share the same average infection probability, independent of their \textcolor{black}{location in the network neighborhood}.
However, due to the localized nature of the infection process, it is more likely that a node will be infected if its neighbor is infected as well. This significantly affects the predicted trajectory of a disease for low average node degree, $k_0$, leading to inaccurate forecasting \textcolor{black}{using the mean-field approximation (dashed black traces in  Fig.~\ref{fig:RRG_Vailidity}).}

The same limitations of mean-field theory appear in network structures with non-random connectivity, such as lattice networks\textcolor{black}{, where nodes are arranged in a highly ordered geometric grid and connect only to their immediate spatial neighbors}. For this reason, many applications require refining mean-field approaches to achieve more reliable forecasting. Previous extensions to mean-field theory, especially in the context of birth-death-movement models, have established a strong foundation for accurately predicting population dynamics on lattices by systematically including spatial correlations into the analysis~\cite{baker2010correcting, markham2013simplified}. 
In these works, each lattice site can be either occupied ($A$) or empty ($\emptyset$). The model permits three processes: birth, in which a cell reproduces into an adjacent empty site, $A+\emptyset \to A+A$; death, in which a cell is removed, leaving an empty site, $A \to \emptyset$; and movement, in which a cell moves to a neighboring empty site, $A\!+\!\emptyset \!\to\! \emptyset \!+\! A$. The birth and death processes map directly onto infection, $I+S \to I+I$, and recovery, $I \to S$, in SIS dynamics, while movement is neglected as we aim to derive corrections to mean-field theory for SIS processes.

The infection process in network-based SIS models relies on transmissions occurring only across edges connecting susceptible and infected individuals. This local interaction structure makes it advantageous to formulate the dynamics directly in terms of the density of connected susceptible-infected pairs, $x_{SI}^{(1)}$, rather than only global densities~\cite{keeling_effects_1999}. Henceforth, the superscript $1$ denotes (distance-one) adjacent lattice sites. Thus, the evolution of $x_{I}$ satisfies
\begin{equation}
    \frac{\dd x_{I}}{\dd t} = \beta k_0 x_{SI}^{(1)}- x_{I}. \label{eq:SISPairwise}
\end{equation}
To quantify deviations from mean-field predictions, we introduce pairwise correlation functions $F_{II}^{(1)}\!(t)$ and $F_{SI}^{(1)}\!(t)$, which measure the actual fractions of infected-infected and susceptible-infected pairs, compared to their mean-field values. More generally, for any network distance $r$, these are defined as
\begin{equation}
    F_{II}^{(r)}(t) = \frac{x_{II}^{(r)}(t)}{x_I(t)^2}, \quad F_{SI}^{(r)}(t) = \frac{x_{SI}^{(r)}(t)}{x_I(t) (1 - x_I(t))}, \label{eq:corrDef}
\end{equation}
where $x_{II}^{(r)}(t)$ and $x_{SI}^{(r)}(t)$ denote the fractions of infected-infected and susceptible-infected pairs at a \textit{distance} $r$ from one another, respectively. The parameter $r$ can refer either to Euclidean distance or shortest path length on the network, but does not explicitly enter the system dynamics, as demonstrated in the following section. The normalization by mean-field edge densities, which assume statistical independence \textcolor{black}{(i.e., that the infection status of one node does not influence the probability of another being infected)}, ensures that
$F_{II}^{(r)}$ and $F_{SI}^{(r)}$ quantify the degree of spatial correlation between node states at distance $r$ on the network. 

Treating $x_{SI}$ and $x_{II}$ as edge probabilities, one has $x_{II}^{(r)} + x_{SI}^{(r)} = x_I^{(r)}= x_I$. This is since $x_I^{(r)}$ is the marginal probability of a node at distance $r$ from the reference node (regardless of its infection status) to be infected and, according to the mean-field assumption, is equal to the fraction of infected individuals across the whole network. Hence, the correlation functions are related as
\begin{equation}
   F_{SI}^{(r)}(t) = \frac{x_I(t) - x_{II}^{(r)}(t)}{x_I(t)(1 - x_I(t))} = \frac{1 - F_{II}^{(r)}(t)x_I(t)}{1 - x_I(t)}. \label{eq:F_SI-to-F}
\end{equation}
Plugging Eqs.~(\ref{eq:corrDef}) and (\ref{eq:F_SI-to-F}) into Eq.~\eqref{eq:SISPairwise}, the pairwise model becomes an explicit correction to the mean-field equation,
\begin{equation}
    \frac{\dd x_{I}}{\dd t} = \beta k_0 x_{I}\left[1-F_{II}^{(1)}x_{I}\right]- x_{I}. \label{eq:SISLogisticCorrectred}
\end{equation}
This equation, valid for homogeneous network topologies, illustrates how the local clustering of infectiousness, as captured by $F_{II}^{(1)}$, alters the prediction of the mean-field theory for the epidemic dynamics.
 
The infected--infected pairwise correlation function, $F_{II}^{(r)}$, approaches unity when the infection status of nodes at distance $r$ are statistically independent, which occurs in well-mixed networks ($k_0 \gg 1$), where infected nodes are expected to be randomly distributed without pronounced spatial clustering. Conversely, $F_{II}^{(r)}$ is greater than unity when infected nodes exhibit clustering (positive spatial correlation), as commonly seen in network-based SIS processes due to local infection spread from neighbors, making infected nodes closer together than expected under independence. Under typical SIS dynamics on connected networks, values of $F_{II}^{(r)} < 1$ do not occur, since infection propagation inherently generates clustering rather than dispersion.  Therefore, the term $1 - F_{II}^{(1)}\!x_I\;$ may be interpreted as an effective susceptible density, and it is typically lower compared to the naive mean-field approximation. Similarly, the behavior of $F_{SI}^{(r)}$ tends to lie below unity due to local depletion of susceptible neighbors around infected sites. \textcolor{black}{This reduction reflects the rapid exhaustion of local susceptible pools, which suppresses the effective transmission rate beyond what the mean-field susceptible density would predict.}
\textcolor{black}{Notably, although the SPM relies on an equation equivalent to Eq.~\eqref{eq:SISLogisticCorrectred} to describe infection dynamics, its formulation for the evolution of $F_{II}^{(1)}$ differs fundamentally (see Sec.~\ref{sec:Dynamics-F}). Consequently, as shown in Fig.~\ref{fig:RRG_Vailidity}, the SPM systematically overestimates the infection density at low average degrees $k_0$, whereas our MPM agrees well with corresponding numerical simulations. }

\vspace{-1mm}
\section{Dependence of nearest-neighbor correlation on network topology}
\label{topology}
\vspace{-1mm}
We now further motivate the development of bespoke corrections to mean-field theory for the RRG by using numerical simulations to highlight significant variations in $F_{II}^{(1)}$, and hence global infection dynamics, with the underlying topology of regular networks \textcolor{black}{(i.e., networks with a fixed node degree $k_0$). To do this, we investigate the effects of rewiring by taking initially two- and three-dimensional regular lattices, which possess a clear spatial embedding and strict local neighborhoods, and randomly exchange their edges with probability $p$. 
This approach, inspired by the methodology of Watts and Strogatz~\cite{watts1998collective},  constructs networks that preserve the fixed node degree but become increasingly random, progressively destroying the original spatial structure as the probability of two spatially adjacent nodes remaining connected decreases with $p$.} The edge swapping algorithm we use is a modification of an algorithm from Wang et al.~\cite{wang2007optimal} and is presented in Appendix A. 
It yields randomized network structures that converge to the RRG as $p$ becomes ${\cal O}(1)$. 

\begin{figure}
    \centering
    \includegraphics[width=0.9\linewidth]{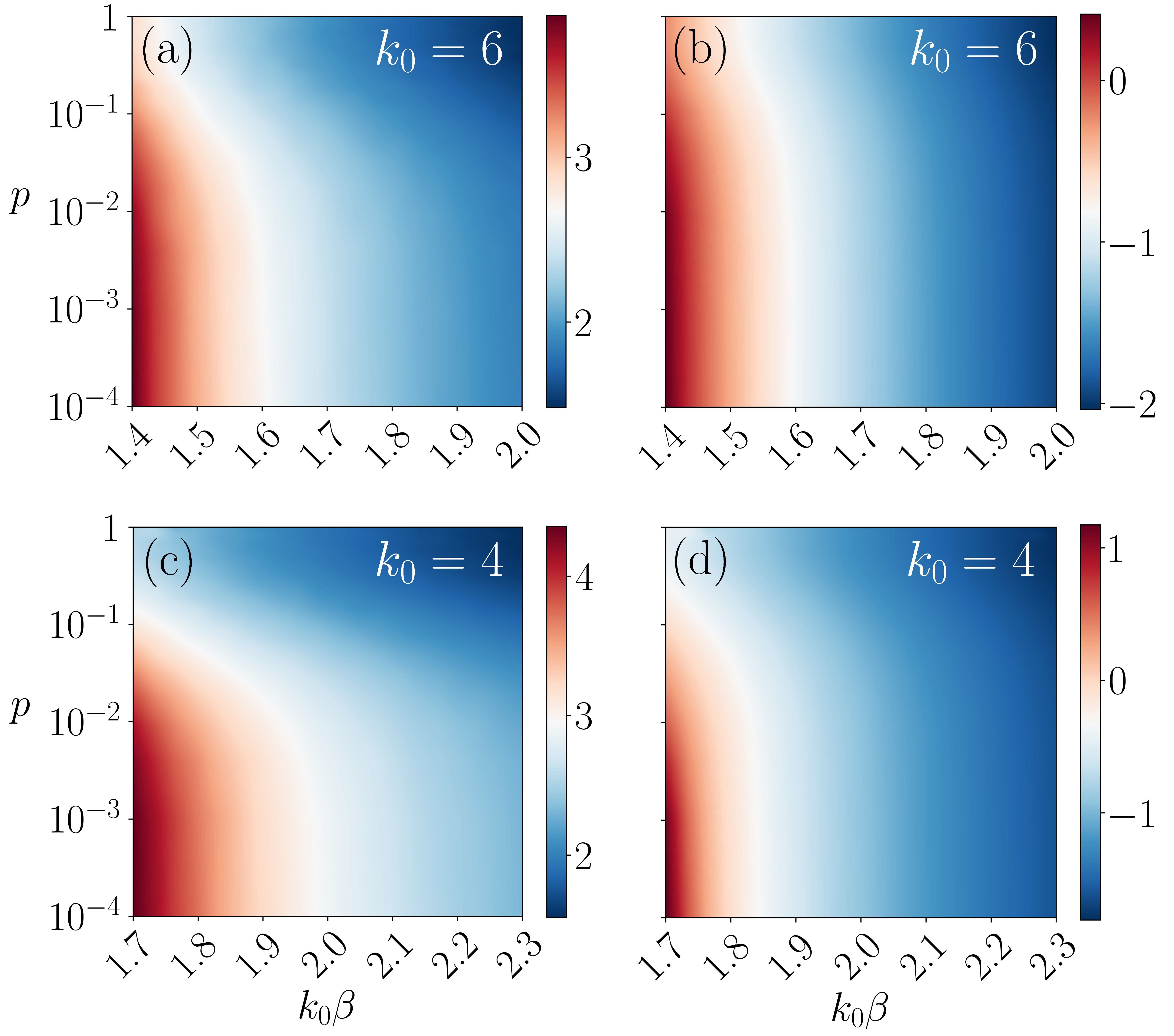}
    \vspace{-5mm}\caption{Relaxation time and steady-state values for the SIS process versus $k_0\beta$ and $p$. Panels (a) and (c) show the natural logarithm of the relaxation times, defined as the time between the peak ($t_{\mathrm{peak}}$) and the first subsequent time $t_{\mathrm{th}}$ at which $F^{(1)}_{II}(t)$ has decayed to $0.1$ above its steady-state value $\bar F^{(1)}_{II}$, whilst (b) and (d) show $\ln(\bar F^{(1)}_{II} - 1)$ for $k_0 = 6$ and $4$, respectively. This parameter sweep shows that increasing $p$ and increasing $k_0\beta$ both reduce the steady-state value of $F^{(1)}_{II}$ and the relaxation time of the process. \textcolor{black}{Simulations were executed on networks with $N \sim 10^4$ nodes (with the exact value depending on the lattice dimension) and $\gamma = 1$. In all simulations, $0.01N$ infectious individuals are randomly placed on the network, with all other nodes occupied by susceptible individuals, and results are averaged over 100 simulations.}}
    \label{fig:paramSweep}
\end{figure}

\textcolor{black}{
To study how increasing topological randomness affects spatial correlations between adjacent nodes and infection trajectories, we perform a parameter sweep over the per-node infection rate, $k_0\beta$, and the edge-swapping probability, $p$, on two- and three-dimensional square lattices, with $k_0=4$ and $k_0=6$, respectively. We find that increasing either $p$ (thereby reducing the mean shortest path length) or $k_0\beta$ substantially suppresses spatial correlations. This is reflected in shorter relaxation times (Figs.~\ref{fig:paramSweep}(a) and \ref{fig:paramSweep}(c)), defined as $\tau_{\mathrm{relax}} = t_{\mathrm{th}} - t_{\mathrm{peak}}$. Here,  $t_{\mathrm{peak}}$ is the time at which $F^{(1)}_{II}$ reaches its maximum, $F^{(1)}_{II,\mathrm{peak}}$, $t_{\mathrm{th}}$ is the time at which it decays to within 10\% of its steady-state value, i.e., $F^{(1)}_{II}(t_{\mathrm{th}}) = \bar F^{(1)}_{II} + 0.1(F^{(1)}_{II,\mathrm{peak}} - \bar F^{(1)}_{II})$, while $\bar F^{(1)}_{II}$ denotes the steady-state value estimated from the long-time tail. The suppression is also evident in the reduced steady-state values of $F_{II}^{(1)}$ (Figs.~\ref{fig:paramSweep}(b) and \ref{fig:paramSweep}(d)). Overall, adding topological randomness increases the steady-state infection density and accelerates the dynamics, making its effect qualitatively similar to that of increasing the infection rate $\beta$. Similar sensitivities to network topology have been reported in the proportion of recovered individuals in related epidemic models~\cite{eames2008modelling, keeling_effects_1999}.}

Notably, the results in Fig.~\ref{fig:paramSweep} suggest that the corrections to mean-field theory proposed for the regular lattice, where $p=0$~\cite{baker2010correcting, markham2013simplified}, are not appropriate for  RRGs, where $p={\cal O}(1)$, because of the strong dependence of $F_{II}^{(1)}$ on $p$. Below, we therefore derive a  correction to mean-field theory based on random network theory, which outperforms both the standard mean-field theory and SPM.

\section{Dynamics of the Pairwise Correlation Function}\label{sec:Dynamics-F}
\begin{figure}
    \centering
    \includegraphics[width=\linewidth]{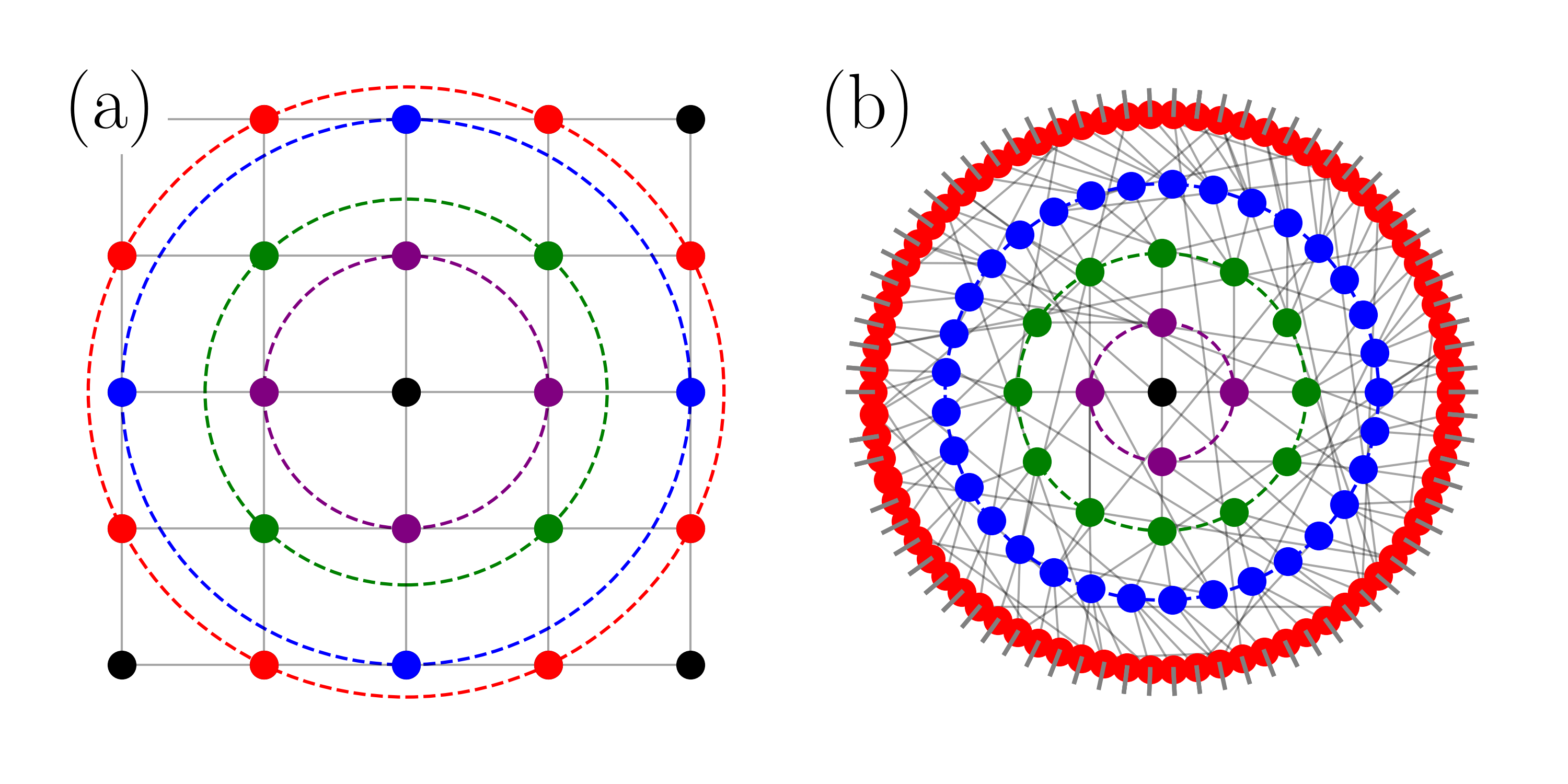}
    \vspace{-11mm}
    \caption{(a) A two-dimensional regular lattice with its origin at $(0,0)$. The purple, green, blue, and red circles represent the first, second, third, and fourth shells, respectively. (b) RRG shells with node degree $k_0=4$. The central node is black, \textcolor{black}{and the shells follow the same color scheme as in (a).}}
    \label{fig:Explaining the grid}
\end{figure}

We now formulate generalized dynamics for the distance-one pairwise correlation function, $F_{II}^{(1)}\left(t\right)$, for both the lattice and the RRG, using a series of hierarchical ODEs for equivalent correlation functions at a range of distances on the network, $\{F_{II}^{(r)}\left(t\right)\}^{r=r_{\text{max}}}_{r=1}$.
On the RRG, which lacks a spatial embedding, the notion of Euclidean distance is ill-defined. 
As such, it is often preferable to consider the shortest path between nodes as a metric of distance. This notion of distance affords the definition of ``shells" of nodes surrounding each node on a network, where shells of distance one contain all nodes directly connected to a given node, shells of distance two contain all nodes connected to the nodes of distance one that are not in shells of distance one, etc. (see Fig.~\ref{fig:Explaining the grid}). 
This, in turn, means that a set of correlation functions $\{F_{II}^{(\ell)}\left(t\right)\}^{\ell=\ell_{\text{max}}}_{\ell=1}$ can be defined in a manner of Eq.~\eqref{eq:corrDef}, where $\ell$ denotes the shell, and \textcolor{black}{ $\ell_{\max}$ is the maximum shell index, defined explicitly in Sec.~\ref{RRGCorrectionsSection}}. In Eqs.~\eqref{eq:corrDef}-\eqref{eq:F_SI-to-F}, the Euclidean distance $r$ appears only in the enumeration of the correlation function and is not explicitly involved in the system dynamics. 
\textcolor{black}{Thus, $r$ can be seamlessly replaced by the shell index $\ell$; e.g., on the two-dimensional lattice, the distances $r=1,\sqrt{2}/2,\sqrt{3}/2,\dots$, map directly to $\ell=1,2,3,\dots$ (see Fig.~\ref{fig:Explaining the grid}). Henceforth, we formulate the dynamics entirely in terms of this shell index.}

The dynamics of the pairwise correlation function $F_{II}^{(\ell)}=x_{II}^{(\ell)}/x_I^2$ can be derived by analyzing the change in both the two-point probability $x_{II}^{(\ell)}$ and global infected density $x_I$. The former includes contributions from infection events in which a susceptible node in an $I$-$S$ pair becomes infected through a third, infected neighbor distinct from the original $I$ node. This process is described by the three-point function $x_{ISI}^{(\ell \ell^\prime m)}$, evaluated via the Kirkwood superposition approximation (KSA). The KSA expresses $x_{ISI}^{(\ell \ell^\prime m)}$ as a product of pairwise distributions,
\begin{equation}
x_{ISI}^{(\ell \ell^\prime m)} \!=\! \frac{x_{SI}^{(\ell)}x_{SI}^{(m)}x_{II}^{(\ell^\prime)}}{(1\!-\!x_I) x_I^2} \!=\! (1 - x_I)x_I^2 F_{SI}^{(\ell)} F_{SI}^{(m)} F_{II}^{(\ell^\prime)}, \label{KSA-SIS}
\end{equation}
where $\ell$, $\ell^\prime$, and $m$ are the shell distances between nodes 1-2, 1-3, and 2-3 of the triplet, respectively. 
\textcolor{black}{This closure differs from that used in the SPM, where the triplet closure is applied only to nearest-neighbor triples (with edges connecting nodes 1-2 and 2-3, yielding $m=\ell=1$), and correlations beyond nearest neighbors are ignored by setting $F_{II}^{(\ell^\prime)}\!\!=\! x_{II}^{(\ell^\prime)} / x_I^2 \!=\! 1$ for all $\ell^\prime > 1$. Thus, Eq.~\eqref{KSA-SIS} is retained only when there is an edge connecting nodes 1-3 ($\ell^\prime = 1$), forming a triangle 1-2-3. Yet, on both the square lattice and RRG, the probability of such triangles vanishes for network size $N\gg 1$. Hence, the SPM reduces to a closure $x_{ISI} = x_{SI}^2/(1 - x_I)$,} \textcolor{black}{neglecting longer-range spatial correlations.}

\textcolor{black}{We now outline the derivation of the dynamics of $F_{II}^{(\ell)}$.} 
\textcolor{black}{To account for this third-party infection process,} we set $m=1$ and sum over all possible third-party shells $\ell^\prime$, weighted by $p(\ell^\prime|\ell)$, the conditional probability that a neighbor of a susceptible node in shell $\ell$ lies in shell $\ell^\prime$. The total contribution to $\dd F_{II}^{(\ell)}/\dd t$ is then equal to
\textcolor{black}{
$2\beta (1 - x_I) F_{SI}^{(\ell)}(t) \sum_{\ell^\prime > 0} k_0 p(\ell^\prime|\ell)\,F_{II}^{(\ell^\prime)}(t)$}, where the factor of two accounts for symmetric $S$-$I$ and $I$-$S$ pairs.  

Another contribution to the dynamics of  $x_{II}^{(\ell)}$ arises from the direct infection of adjacent $I$-$S$ pairs, which converts them into $I$-$I$ pairs. This occurs only when $\ell=1$, and can thus be written as $ 2\beta\delta_{\ell,1}x_{SI}^{(1)}/x_I^2=2\beta\delta_{\ell,1} \left(1-x_{I} \right)F_{SI}^{(1)}/x_I$, where the factor of two follows similarly as before, and $\delta_{n,m}$ is the Kronecker delta.   
Furthermore, $F_{II}^{(\ell)}$ is also affected by variations in the overall infection density $x_I$ that appear in the denominator of $F_{II}^{(\ell)}$. Differentiating $F_{II}^{(\ell)} = x_{II}^{(\ell)}/x_I^2$ yields a term $-2 x_{II}^{(\ell)} (\dd x_I/\dd t) / x_I^3$, which equals $-2 F_{II}^{(\ell)} (\dd x_I/\dd t) / x_I$. Substituting the infectious term from Eq.~\eqref{eq:SISLogisticCorrectred} and using the identity $(1-x_I)F_{SI}^{(\ell)} = 1 - x_I F_{II}^{(\ell)}$ from Eq.~\eqref{eq:F_SI-to-F} recasts this contribution as $-2 F_{II}^{(\ell)}\,\beta k_0 (1-x_I)F_{SI}^{(1)}$. Here, the factor of two reflects the squared normalization in $x_I$. 

Finally, while recovery events reduce both $x_{II}^{(\ell)}$ and $x_I$, their effects cancel out in the evolution of $F_{II}^{(\ell)}\left(t\right)$. \textcolor{black}{Specifically, recovery of either node in an $I$-$I$ pair decreases the numerator $x_{II}^{(\ell)}$ at a rate $2 x_{II}^{(\ell)}$, contributing $-2 F_{II}^{(\ell)}$ to the time derivative. Simultaneously, \textcolor{black}{the recovery term in Eq.~\eqref{eq:SISLogisticCorrectred} acts through the derivative of the $x_I^2$ denominator, $-2 F_{II}^{(\ell)} (\dd x_I/\dd t) / x_I$, yielding} an exactly compensating $+2 F_{II}^{(\ell)}$ term, resulting in a net zero effect. }

In total, the dynamics of $F_{II}^{(\ell)}$ are described by the coupled system of ODEs
\begin{align}
    \frac{\dd F_{II}^{(\ell)}}{\dd t} &= \left[F_{SI}^{(\ell)}  \sum_{\ell^{\prime}>0}p(\ell^{\prime}|\ell) F_{II}^{(\ell^{\prime})} \!+\! \frac{\delta_{\ell,1}}{k_0 x_{I}} \!-\!F_{II}^{(\ell)} \right]  \nonumber \\ 
    &\hphantom{=[} \times \vphantom{\frac{1-F_\ell\left(t\right)x_{I}}{1-x_{I}}} 2\beta k_0 \left(1-x_{I} \right)F_{SI}^{(1)}, \label{eq:corrDyn}
\end{align}
where $\ell \in \{1,\dots,\ell_{\max} \}$, and $p(m|n)$ is the conditional probability that a given neighbor of a node in shell $n$ lies in shell $m$. Note that, the dynamics can also be written entirely in terms of $F_{II}^{(\ell)}$ by using Eq.~\eqref{eq:F_SI-to-F}, which is convenient for numerical solutions. 

Importantly, 
Eq.~\eqref{eq:corrDyn} applies both to the regular lattice and to the RRG.
For example, in a two‑dimensional square lattice (Fig.~\ref{fig:Explaining the grid}(a)), a node in \((1,0)\) (the first shell) has one neighbor in the zeroth shell, two in the second shell, and one in the third shell, such that $p(k|1) = \delta_{k,0}/4 + \delta_{k,2}/2 + \delta_{k,3}/4$.
Similarly, a node in \((-1,1)\) (the second shell) has two neighbors in the first shell and two in the fourth shell, giving $p(k|2) = \delta_{k,1}/2 + \delta_{k,4}/2$.
By translational invariance, the probabilities \(p(\ell^\prime|\ell)\) are uniform across nodes, and hence are global constants that vary with the lattice dimension.

\section{Computing Shell Probabilities on Random Regular  Graphs}
\label{RRGCorrectionsSection}

To complete the formulation of the dynamics of the pairwise correlation function, Eq.~(\ref{eq:corrDyn}), it remains to find an explicit expression for  $p(\ell^\prime|\ell)$ in the case of the RRG. For the latter, the fraction of individuals located in shells beyond distance $\ell$ follows the Gompertz distribution~\cite{Nitzan2016}
\begin{equation}
P\left(L>\ell\right)=\exp\left\{ -\frac{k_0} {\left(k_{0}-2\right)N} \left(\left(k_{0}-1\right)^{\ell}-1\right)\right\}, \label{eq:p_gompertz}
\end{equation}
where $P\left(L>\ell\right)$ is the probability that a randomly selected node on the network lies at a greater distance than $\ell$ from a given node. 
The difference between the complementary cumulative fractions of consecutive shells, therefore, determines the fraction of nodes in a specific shell $ p\left(\ell\right)=P\left(L>\ell-1\right)-P\left(L>\ell\right)$,
and the number of nodes in the $\ell$-th shell is thus given by
\begin{equation}
    N_{\ell}=Np\left(\ell\right)=N\left[P\left(L>\ell-1\right)-P\left(L>\ell\right)\right]. \label{eq: number of nodes in the shell}
\end{equation}

The conditional probability $p(k|n)$ quantifies the likelihood that a randomly chosen neighbor of a node in the $n$-th shell belongs to the $k$-th shell, and plays a crucial role in characterizing network connectivity at increasing distances from a reference node. At this point, we note that the number of nodes in the $\ell$-th shell satisfies
\begin{equation}
    N_{\ell} = \prod_{k=0}^{\ell-1} \frac{p(k+1\,|\,k)}{p(k\,|\,k+1)},
\end{equation}
where this identity stems from Bayes' theorem
\begin{equation}
    p(k|k+1) = p(k+1|k) \frac{p(k)}{p(k+1)} 
    = p(k+1|k) \frac{N_k}{N_{k+1}},
    \label{eq: p^s bayes relation}
\end{equation}
with $p(k)$ being proportional to the shell population $N_k$. 

\textcolor{black}{The calculation of the conditional probabilities $p(\ell-1|\ell)$, $p(\ell|\ell)$, and $p(\ell+1|\ell)$ for a target node in the $\ell$-th shell is performed iteratively. At each step, the recursion exploits the previously obtained $p(\ell|\ell-1)$ and the shell populations $N_0,\dots,N_\ell$ to advance to the next shell. From Bayes' theorem (Eq.~\eqref{eq: p^s bayes relation}), the backward probability is simply $p\left(\ell-1|\ell\right)=p\left(\ell|\ell-1\right)N_{\ell-1}/N_{\ell}$. Next, the probability $p(\ell|\ell)$ describes the likelihood that a neighbor of a node in shell $\ell$ also lies in the same shell. We compute this using a combinatorial stub-matching argument. The number of available ``stubs'' (half-edges yet to be paired to form edges) in shell $\ell$ that can connect internally is the total capacity $k_0 N_\ell$, minus the stubs already paired to form edges to the previous shell $\ell-1$, and excluding trivial self-connections. Dividing this by the total remaining available stubs in the network (after accounting for all connections in shells $0, \dots, \ell$) and simplifying the resulting algebraic expression, yields:
\begin{equation}
    p(\ell|\ell) = \frac{ (N_\ell - 1)\, \left[ 1 - p(\ell-1|\ell) \right]^2 }
    { N - \sum\limits_{k=0}^{\ell} N_k + (N_\ell - 1)\left[ 1 - p(\ell-1|\ell) \right] }.\label{eq:p_l given l}
\end{equation}
This expression captures the self-closing probability within the shell as the ratio of available internal same-shell connections to all possible remaining connections.}

\textcolor{black}{The quantity $p(\ell+1|\ell)$, representing the probability that a neighbor of a node in shell $\ell$ resides in the forward shell $\ell+1$, can be formulated either by enforcing the complementary probability, $p(\ell+1|\ell)=1-p(\ell|\ell)-p(\ell-1|\ell)$, or by employing the same combinatorial logic where the numerator corresponds to available endpoints in the forward shells and the denominator accounts for all remaining stubs. The system is rendered self-consistent by stipulating the boundary condition $p(1|0) = 1$, ensuring that the initial shell is reached deterministically from the origin. With this assignment, the recursive relations for the shell populations $N_k$ and the transition probabilities $p(k|\ell)$ are fully formulated, thereby defining the complete hierarchy of shell-wise network topology.}

\textcolor{black}{To assess the accuracy of our MPM against the SPM and mean-field theories on the RRG, we directly compare analytical solutions to numerical simulations of the SIS model. Specifically, we solve a system of coupled ODEs for the pairwise correlation functions in each shell $\ell$ (Eq.~\eqref{eq:corrDyn}), truncated at a maximum shell index $\ell_{\max}$. We dynamically define $\ell_{\max}$ as the smallest shell index for which the expected number of nodes in the next shell falls below unity ($N_{\ell_{\max}+1} < 1$), determined via Eq.~\eqref{eq: number of nodes in the shell}. This ensures that shells beyond $\ell_{\max}$ contain a negligible fraction of the network and thus, contribute minimally to the correlation dynamics. For the network sizes considered ($N = 10^4$), this yields $\ell_{\max} \approx 5$--$9$ for $k_0 = 3$--$10$ (where $\ell_{\max}$ decreases with increasing $k_0$), capturing the essential spatial structure while remaining computationally tractable. The conditional probabilities determine how correlations propagate through the network topology, ensuring consistency in the shell structure. Once solved, the hierarchy of correlation functions, particularly the distance-one correlation function, is then used in Eq.~\eqref{eq:SISLogisticCorrectred} to predict the global infection density. }

This unified approach enables direct comparison between theory and simulation. We illustrate this on a network of $N=10^4$ nodes, with an initial fraction $x_I(0)=10^{-2}$ of infected individuals.
\textcolor{black}{The results shown in Fig.~\ref{fig:RRG_Vailidity} display good agreement between the numerical solution of Eq.~\eqref{eq:corrDyn} (blue) and discrete simulations (red). Notably, the MPM systematically outperforms the SPM (dark green), which overestimates the steady-state infection density and underestimates the correlation function $F_{II}^{(1)}$, especially at lower node degrees.} Note that, the slight discrepancy between the MPM and simulations is due to the proximity of $\beta$ to the critical infection rate: $\beta_c k_0/\gamma=k_0/(k_0-1)$ for the RRG~\cite{morita_basic_2022}; \textcolor{black}{the latter separates the endemic regime, $\beta > \beta_c$ (persistence of infection), from the absorbing regime, $\beta < \beta_c$ (epidemic dies out deterministically). }Close to the critical points ($\beta_c k_0/\gamma=1.5$ and $1.33$ for $k_0=3$ and $4$, respectively) the epidemic dynamics are dominated by strong fluctuations and long-range correlations, and are inherently more difficult to capture in low-order moment closure, such as the KSA~\cite{baker2010correcting, markham2013simplified}.
As was shown (see also below), the agreement between  model simulations and our MPM improves with \textcolor{black}{increasing node degree $k_0$ or infection rate $\beta$.}

\section{A modified steady-state analysis on random regular networks}
Alongside the dynamical trajectory of infection density, we now compute the steady-state value of infection density, $\bar{x}_{I}$, predicted within the MPM framework. This quantity is of particular interest in percolation theory, where the non-equilibrium phase transition between endemicity and disease elimination at a critical infection rate, $\beta_{c}$, is not fully understood on a wide range of network topologies. 
Setting $\dd x_{I} / \dd t = 0$ in Eq.~\eqref{eq:SISLogisticCorrectred}, we find
\begin{equation}
    \bar{x}_{I} = \bar{x}_{I}^{\text{MF}}/\bar{F}_{II}^{(1)}, \label{eq: x_{I} steady-state}
\end{equation}
where $\bar{x}_{I}^{\text{MF}}= 1 - \gamma/(\beta k_0)\equiv 1-1/\Lambda$ is the standard mean-field prediction for the endemic state, obtained by finding the steady-state value of Eq.~\eqref{eq:SISLogisticCorrectred} with $F_{II}^{(1)}(t)=1$, while $\bar{F}_{II}^{(1)}$ is the steady-state value of the distance-one pairwise correlation function. \textcolor{black}{The ratio $\Lambda = \beta k_0/\gamma$ is known as the effective spreading rate~\cite{PastorSatorras2001, PastorSatorras2015}, a dimensionless control parameter that dictates the epidemic threshold $\Lambda=1$ in the well-mixed limit. }

\begin{figure}
    \centering
    \includegraphics[width=0.87\linewidth]{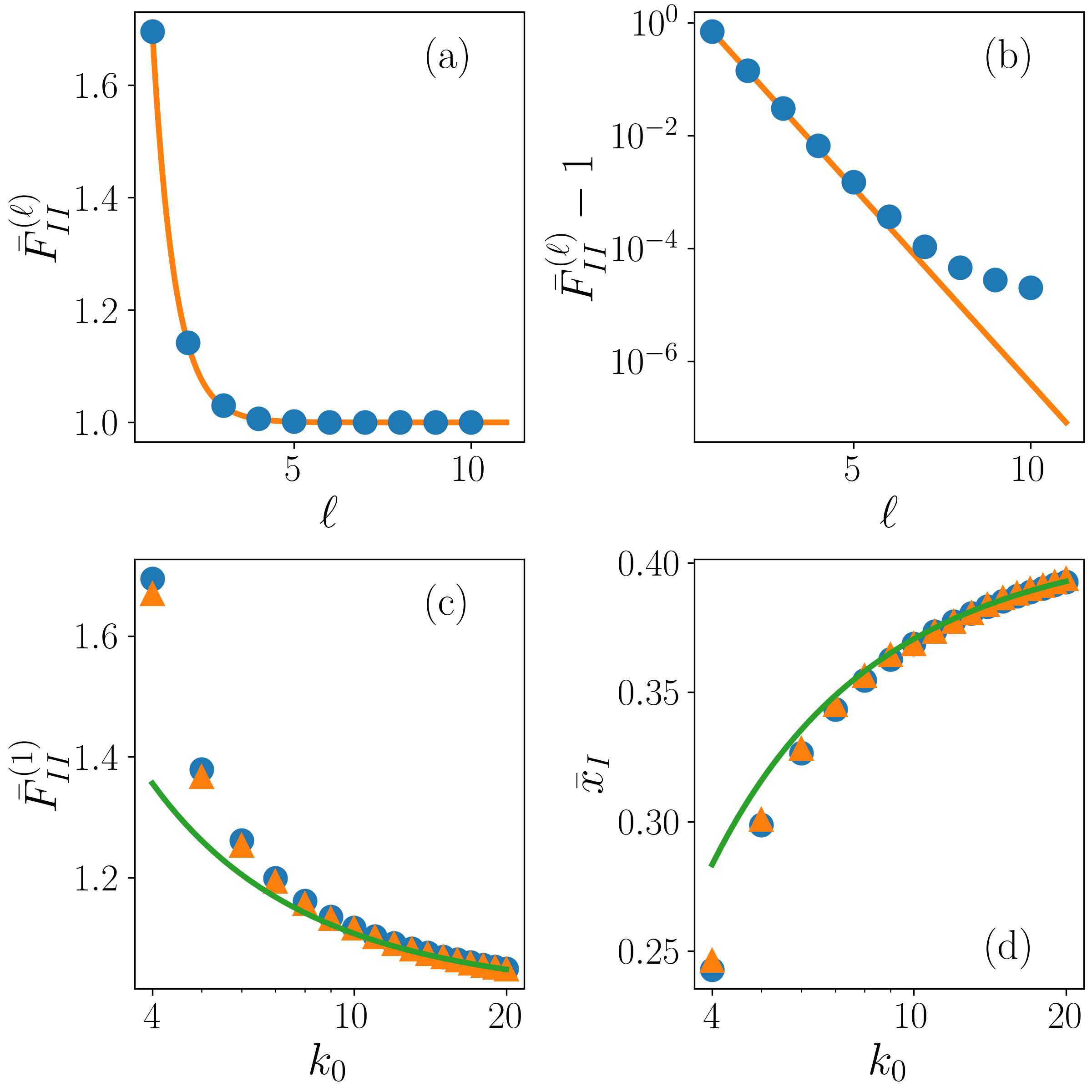}
    \vspace{-5mm}\caption{(a)-(b) The steady-state pairwise correlation function, $\bar{F}_{II}^{(\ell)}$, versus the network distance $\ell$, on linear and semi-logarithmic scales, respectively, for $\Lambda=1.7$. Circles indicate values obtained from numerical simulations, while solid lines are exponential fits based on Eq.~\eqref{eq: F_ss}. (c)-(d)  $\bar{F}_{II}^{(1)}$ and  $\bar{x}_{I}$, respectively, versus the average degree $k_0$: comparison between numerical solutions (circles), simulation results (triangles), and the large-$k_0$ asymptotes (solid green lines), see text.}
    \label{fig:ExponentialFitSteadyState}
\end{figure}

For an RRG, $\bar{F}_{II}^{(\ell)}$ is obtained by setting $\dd F_{II}^{(\ell)} / \dd t = 0$ in Eq.~\eqref{eq:corrDyn} and using the expression for $\bar{x}_{I}^{\text{MF}}$, yielding
\begin{equation}
    \hspace{-0.3cm}\bar{F}_{II}^{(\ell)} \!=\! \frac{\bar{F}_{II}^{(1)}\!-\!\bar{F}_{II}^{(\ell)}\bar{x}_{I}^{\text{MF}}}{\bar{F}_{II}^{(1)}-\bar{x}_{I}^{\text{MF}}}\!\!\!\!\!\!\!\sum_{\ell^\prime=\max\{\ell-1,1\}}^{\ell+1}\!\!\!\!\!\!\!\!\!\!\!\!p\left(\ell^\prime|\ell\right)\bar{F}_{II}^{(\ell^\prime)} 
    + \frac{\bar{F}_{II}^{(1)}\delta_{\ell,1}}{k_0 \bar{x}_{I}^{\text{MF}}}. \label{eq: F_ii ss}
\end{equation}
Analysis of the steady-state behavior of the pairwise correlation function on the RRG shows an exponential decay of $\bar{F}_{II}^{(\ell)}$ with increasing distance between nodes (Figs.~\ref{fig:ExponentialFitSteadyState}(a)-(b)). This decay is accurately described by
\begin{equation}
\bar{F}_{II}^{(\ell)} =  1 + \alpha e^{-\lambda\ell} \equiv 1 + \alpha K^\ell,
\label{eq: F_ss}
\end{equation}
where $\alpha$ and $K=\exp(-\lambda)$ are parameters yet to be determined. This empirical ansatz allows one to approximate $\bar{F}_{II}^{(\ell)}$. 
Substituting Eq.~\eqref{eq: F_ss} into~\eqref{eq: F_ii ss} for shells $\ell=1,2$ and further using Eq.~(\ref{eq: F_ss}) for $\ell=3$, yields two implicit equations for $\alpha$ and $K$. 
Following algebraic manipulations detailed in Appendix~\ref{appendix-ss}, these equations reduce to a cubic equation $AK^3 + BK^2 + CK + D=0$, alongside an explicit equation connecting $\alpha$ and $K$, which can be solved numerically for any $k_0$. Notably, while Eq.~(\ref{eq: F_ii ss}) breaks down at larger values of $\ell$ (see Fig.~\ref{fig:ExponentialFitSteadyState}(b)), it remains rather accurate for $\ell={\cal O}(1)$; this allows for an accurate approximation of the distance-one pairwise correlation, which is the primary quantity of interest in our analysis, as it determines the correction to  $\bar{x}_I$. 

The analysis above greatly simplifies under the assumption of $k_0 \ll N^{1/2}$, making the network locally tree-like~\footnote{In the complementary regime $k_0 \gtrsim N^{1/2}$, the RRG is no longer locally tree-like and finite-size effects as well as an increased prevalence of short cycles in the network occurs. Yet, as the expected number of shells scales as $\log_{k_0}\!N$, for $k_0\gtrsim N^{1/2}$ the network has only $\sim 2$ shells, which is clearly a less interesting regime to deal with.}. In this limit, the conditional probabilities read
\begin{eqnarray}
&p(0|0)\!=\!0, \;\; p(1|0)\!=\!1, \;\;
p(0|1)\!=\!1/k_0,\;\;p(1|1)=\mathcal{O}(k_0/N), \nonumber \\ 
& p(2|1)\!=\!1 - 1/k_0 + \mathcal{O}(k_0/N),\;\;p(1|2)\!=\!1/k_0 + \mathcal{O}(k_0/N), \nonumber \\
&\hspace{-12mm}p(2|2)\!=\! \mathcal{O}(k_0^2/N), \;\;
p(3|2)\!=\!1 - 1/k_0 + \mathcal{O}(k_0^2/N).
\end{eqnarray}
Substituting these results and Eq.~\eqref{eq: F_ss} into~\eqref{eq: F_ii ss}, yields
\begin{eqnarray}
    &\alpha =-\left[\left(k_{0}-1\right)\left(\Lambda-1\right)K\left(K-\Lambda\right)+\Lambda K\right]^{-1}\label{eq: steady-state begin small k0}, \nonumber\\
    &0 = K^{3}-\Lambda\frac{\left(2k_{0}-1\right)}{k_{0}}K^{2}+\left[\Lambda^{2}-\frac{\Lambda-1}{k_{0}-1}\right]K-\frac{\Lambda}{k_{0}}\label{eq: steady-state end small k0}.
\end{eqnarray}
Further assuming  $k_0\gg 1$, and expanding $K$ and $\alpha$ in powers of $k_0^{-1}$ in the regime $1 \ll k_0 \ll N^{1/2}$ gives
$K = 1/(\Lambda k_0)+(\Lambda + 1)/(\Lambda^3 k_0^2) + \mathcal{O}\left(k_0^{-3}\right)$, and $\alpha = 1/(\Lambda - 1) + (\Lambda^2-\Lambda+1)/[\Lambda(\Lambda - 1)^2 k_0]  + \mathcal{O}\left(k_0^{-2}\right)$.
Here, we have retained terms in the series expansion of $\alpha$ up to ${\cal O}(k_0^{-1})$, since the leading term in the expansion of $K$ is also of ${\cal O}(k_0^{-1})$, while we are interested in expressing their product $\bar{F}_{II}^{(1)} = 1 + \alpha K$  up to ${\cal O}(k_0^{-2})$. This yields
\begin{equation}
    \bar{F}_{II}^{(1)} 
    \!=\! 1 \!+\! \frac{1}{(\Lambda \!-\! 1) \Lambda k_0}
    \!+\! \frac{\Lambda^3 + \Lambda - 1}{(\Lambda \!-\! 1)^2 \Lambda^3 k_0^2} \!+\! \mathcal{O}\!\!\left(\!\frac{1}{k_0^3}\!\right)\!. \label{F-bar-power-series}
\end{equation}
Therefore, using Eq.~(\ref{eq: x_{I} steady-state}), we obtain
\begin{equation}
    \bar{x}_I 
    = \bar{x}_{I}^{\text{MF}} - \frac{1}{k_0\Lambda^2} -  \frac{\Lambda^2+\Lambda+1}{k_0^2\Lambda^4} - \mathcal{O}\left(\frac{1}{k_0^3}\right). \label{x-bar-power-series}
\end{equation}
Our predictions for $\bar{F}_{II}^{(1)}$ and $\bar{x}_I$ agree well with numerical simulations, see Figs.~\ref{fig:ExponentialFitSteadyState}(c)-(d). These expansions make explicit how finite-degree corrections alter both the two-point correlation function and infection density relative to the standard mean-field predictions. The fact that the correction terms in $\bar{x}_I$ are negative reflects the dampening effect of local clustering and correlations on disease spread in finite-connectivity networks, consistent with $\bar{F}_{II}^{(1)}(t) \geq 1$ at all distances, as seen in Fig.~\ref{fig:RRG_Vailidity}. 
\textcolor{black}{Notably, equating Eq.~\eqref{x-bar-power-series} to zero yields the critical spreading rate 
\begin{equation}
    \Lambda_c 
    = 1 + k_0^{-1} + 2k_0^{-2} + \mathcal{O}\left(k_0^{-3}\right), \label{Lambda_c}
\end{equation}
which differs from the SPM threshold at order $\mathcal{O}(k_0^{-2})$, see Appendix C for  a detailed comparison between the MPM and  SPM~\footnote{\textcolor{black}{Note that, once the corrected critical spreading rate 
$\Lambda_c$
 is computed, the basic reproduction number can be readily obtained as 
$R_0=\Lambda/\Lambda_c$.}}.}

Finally, exploiting the power series expansions of the steady-state infection density $\bar{x}_I$ and pairwise correlation function $\bar{F}_{II}^{(1)}$, we also perform a linear stability analysis of the coupled epidemic dynamics given by Eqs.~\eqref{eq:SISLogisticCorrectred} and~\eqref{eq:corrDyn} for $\ell=1$, assuming that the relaxation time of all $\bar{F}_{II}^{(\ell)}$, for $\ell>1$, is comparable to that of $\bar{F}_{II}^{(1)}$. By evaluating the Jacobian matrix at the endemic equilibrium, we explicitly capture corrections beyond the mean-field limit. Expanding the resulting relaxation time (corresponding to the \textcolor{black}{eigenvalue with the smallest magnitude}) up to ${\cal O}(k_0^{-1})$ we find \textcolor{black}{
$\tau_{\mathrm{relax}} \simeq 1/(\Lambda-1)  \left[1 + 2(2 \Lambda-1)/(\left(\Lambda-1\right)^2 k_0)  \right]$.}
The zeroth-order value $1/(\Lambda-1)$ is the mean-field relaxation time, while higher-order positive corrections imply that for lower $k_0$, the system takes more time to converge. 

\section{Discussion} 
In this work, we have characterized, through numerical simulations and analytical calculations, the role of spatial correlations in the transmission dynamics of SIS processes across various network topologies. Our findings demonstrate that spatial correlations between infectious nodes are suppressed both by random edge rewiring, which decreases the average shortest-path length, and by increasing the infection rate. 
The observed sensitivity of spatial correlations to network topology implies that corrections to mean-field theory previously derived for regular lattices are unsuitable for RRGs. To address this, we leveraged existing analytical results on RRG connectivity to develop appropriate mean-field corrections tailored specifically to  random  regular topologies. 
\textcolor{black}{These corrections provide  improved predictions of infection dynamics of SIS processes on RRGs, when the effective spreading rate is above the  epidemic threshold, i.e., $\Lambda>\Lambda_c$. Traditionally, the well-mixed approximation assumes this critical transition occurs at $\Lambda=1$, predicting a steady-state infected fraction $\bar{x}_{I}^{\text{MF}}=1-1/\Lambda$. Our corrections refine this estimate by explicitly accounting for the spatial clustering of infected nodes. Thus, the effective pool of susceptible neighbors per infected individual is smaller than that assumed in mean-field theory, thereby lowering the predicted $\bar{x}_{I}^{\text{MF}}$. Our results are also superior compared to the SPM, especially at low average degrees where multi-shell correlations are most significant (see Appendix~\ref{appendix-c}).}

While the MPM outperforms previous lattice-based corrections in this regime, the accuracy of the predictions still depends on the proximity of the rate of infection, $\beta$, to the critical value for a given underlying network structure. Notably, the accuracy of these corrections diminishes near the critical threshold, indicating the need for more robust approximation methods close to criticality. 

Future research could attempt to re-derive corrections for the RRG by invoking the Fisher-Kopeliovich approximation, which is known to provide improved accuracy where the Kirkwood superposition approximation becomes inadequate~\cite{omelyan2020spatial} (albeit at the expense of greater analytical complexity). 
Another potential direction would be to expand the current framework to heterogeneous topologies. This will enable us to predict and analyze more realistic scenarios, particularly where degree variability plays a crucial role in epidemic outcomes~\cite{hindes_degree_2019}. 

Overall, this work broadens the applicability of existing corrections to mean-field theory and highlights how network structure impacts spatial correlations in SIS processes. 
Another promising future direction involves exploring  network structures with intermediate randomness to bridge the analytical gap between regular lattices and random regular topologies.

\bibliography{bibl}

@article{baker2010correcting,
  title={Correcting mean-field approximations for birth-death-movement processes},
  author={Baker, Ruth E and Simpson, Matthew J},
  journal={Physical Review E—Statistical, Nonlinear, and Soft Matter Physics},
  volume={82},
  number={4},
  pages={041905},
  year={2010},
  publisher={APS}
}

@article{watts1998collective,
  title={Collective dynamics of ‘small-world’ networks},
  author={Watts, Duncan J and Strogatz, Steven H},
  journal={Nature},
  volume={393},
  number={6684},
  pages={440--442},
  year={1998},
  publisher={Nature Publishing Group}
}

@article{wang2007optimal,
  title={Optimal synchronizability of networks},
  author={Wang, B and Zhou, T and Xiu, ZL and Kim, BJ},
  journal={The European Physical Journal B},
  volume={60},
  pages={89--95},
  year={2007},
  publisher={Springer}
}

@article{Nitzan2016,
   author = {Mor Nitzan and Eytan Katzav and Reimer Kühn and Ofer Biham},
   issn = {24700053},
   issue = {6},
   journal = {Physical Review E},
   month = {6},
   publisher = {American Physical Society},
   title = {Distance distribution in configuration-model networks},
   volume = {93},
   year = {2016},
   pages = {062309},
}

@article{PastorSatorras2001,
   author = {Romualdo Pastor-Satorras and Alessandro Vespignani},
   doi = {10.1103/PhysRevLett.86.3200},
   issn = {00319007},
   issue = {14},
   journal = {Physical Review Letters},
   month = {4},
   pages = {3200-3203},
   pmid = {11290142},
   title = {Epidemic spreading in scale-free networks},
   volume = {86},
   year = {2001},
}

@article{kermack1932contributions,
  title={Contributions to the mathematical theory of epidemics. II.—The problem of endemicity},
  author={Kermack, William Ogilvy and McKendrick, Anderson G},
  journal={Proceedings of the Royal Society of London. Series A, containing papers of a mathematical and physical character},
  volume={138},
  number={834},
  pages={55--83},
  year={1932},
  publisher={The Royal Society London}
}

@article{PastorSatorras2015,
   author = {Romualdo Pastor-Satorras and Claudio Castellano and Piet Van Mieghem and Alessandro Vespignani},
   issn = {15390756},
   issue = {3},
   journal = {Reviews of Modern Physics},
   month = {8},
   publisher = {American Physical Society},
   title = {Epidemic processes in complex networks},
   volume = {87},
   year = {2015},
}

@article{da2008global,
  title={Global epidemiology of sexually transmitted diseases},
  author={Da Ros, Carlos T and da Silva Schmitt, Caio},
  journal={Asian Journal of Andrology},
  volume={10},
  number={1},
  pages={110--114},
  year={2008},
  publisher={Wiley Online Library}
}

@article{turner1997epidemiology,
  title={Epidemiology, pathogenesis, and treatment of the common cold},
  author={Turner, Ronald B},
  journal={Annals of Allergy, Asthma \& Immunology},
  volume={78},
  number={6},
  pages={531--540},
  year={1997},
  publisher={Elsevier}
}

@article{markham2013simplified,
  title={Simplified method for including spatial correlations in mean-field approximations},
  author={Markham, Deborah C and Simpson, Matthew J and Baker, Ruth E},
  journal={Physical Review E—Statistical, Nonlinear, and Soft Matter Physics},
  volume={87},
  number={6},
  pages={062702},
  year={2013},
  publisher={APS}
}

@article{seno2020sis,
  title={An SIS model for the epidemic dynamics with two phases of the human day-to-day activity},
  author={Seno, Hiromi},
  journal={Journal of Mathematical Biology},
  volume={80},
  number={7},
  pages={2109--2140},
  year={2020},
  publisher={Springer}
}

@article{real2007spatial,
  title={Spatial dynamics and genetics of infectious diseases on heterogeneous landscapes},
  author={Real, Leslie A and Biek, Roman},
  journal={Journal of the Royal Society Interface},
  volume={4},
  number={16},
  pages={935--948},
  year={2007},
  publisher={The Royal Society London}
}

@article{eames2004monogamous,
  title={Monogamous networks and the spread of sexually transmitted diseases},
  author={Eames, Ken TD and Keeling, Matt J},
  journal={Mathematical Biosciences},
  volume={189},
  number={2},
  pages={115--130},
  year={2004},
  publisher={Elsevier}
}

@article{eames2008modelling,
  title={Modelling disease spread through random and regular contacts in clustered populations},
  author={Eames, Ken TD},
  journal={Theoretical Population Biology},
  volume={73},
  number={1},
  pages={104--111},
  year={2008},
  publisher={Elsevier}
}

@article{omelyan2020spatial,
  title={Spatial population dynamics: beyond the Kirkwood superposition approximation by advancing to the Fisher--Kopeliovich ansatz},
  author={Omelyan, Igor},
  journal={Physica A: Statistical Mechanics and its Applications},
  volume={544},
  pages={123546},
  year={2020},
  publisher={Elsevier}
}

@article{morita_basic_2022,
	title = {Basic reproduction number of epidemic models on sparse networks},
	volume = {106},
	issn = {24700053},
	number = {3},
	journal = {Physical Review E},
	author = {Morita, Satoru},
	month = sep,
	year = {2022},
	pmid = {36266828},
}

@article{hindes_degree_2019,
	title = {Degree {Dispersion} {Increases} the {Rate} of {Rare} {Events} in {Population} {Networks}},
	volume = {123},
	issn = {10797114},
	doi = {10.1103/PhysRevLett.123.068301},
	number = {6},
	journal = {Physical Review Letters},
	author = {Hindes, Jason and Assaf, Michael},
	month = aug,
	year = {2019},
	pmid = {31491193},
	note = {arXiv: 1901.03158
Publisher: American Physical Society},
}

@article{keeling_effects_1999,
	title = {The effects of local spatial structure on epidemiological invasions},
	volume = {266},
	url = {https://royalsocietypublishing.org/doi/10.1098/rspb.1999.0716},
	doi = {10.1098/rspb.1999.0716},
	number = {1421},
	journal = {Proceedings of the Royal Society of London. Series B: Biological Sciences},
	author = {Keeling, M J},
	year = {1999},
	pages = {859--867},
}

@article{keeling_correlation_1997,
	title = {Correlation models for childhood epidemics},
	volume = {264},
	issn = {0962-8452, 1471-2954},
	url = {https://royalsocietypublishing.org/doi/10.1098/rspb.1997.0159},
	doi = {10.1098/rspb.1997.0159},
	number = {1385},
	urldate = {2025-09-05},
	journal = {Proceedings of the Royal Society of London. Series B: Biological Sciences},
	author = {Keeling, M. J. and Rand, D. A. and Morris, A. J.},
	month = aug,
	year = {1997},
	pages = {1149--1156},
}

@article{eames_modeling_2002,
	title = {Modeling dynamic and network heterogeneities in the spread of sexually transmitted diseases},
	volume = {99},
	issn = {0027-8424, 1091-6490},
	url = {https://pnas.org/doi/full/10.1073/pnas.202244299},
	doi = {10.1073/pnas.202244299},
	number = {20},
	urldate = {2025-09-05},
	journal = {Proceedings of the National Academy of Sciences},
	author = {Eames, Ken T. D. and Keeling, Matt J.},
	month = oct,
	year = {2002},
	pages = {13330--13335},
}

@book{kiss2017mathematics,
  title={Mathematics of Epidemics on Networks: From Exact to Approximate Models},
  author={Kiss, Istv{\'a}n Z and Miller, Joel C and Simon, P{\'e}ter L},
  year={2017},
  publisher={Springer},
  address={Cham, Switzerland},
  series={Interdisciplinary Applied Mathematics},
  volume={46},
  isbn={978-3-319-50804-7},
  doi={10.1007/978-3-319-50806-1}
}

\appendix

\section{Edge-swapping algorithm}
Here, we present an edge swapping
algorithm which is a modification of an algorithm from
Wang et al.~\cite{wang2007optimal}. Given a network $G(V, E)$ representing a regular lattice, where $E$ and $V$ are the edge and node sets, the algorithm can be summarized as follows: 
\begin{itemize}   
    \item Sample edges from $E$ randomly without replacement to create a list of all edges in $E$ in random order, say $L=(e_1, e_2,\dots,e_{|E|})$.   
     \item
    Select a variable $p\in [0,1]$, representing the probability of edge exchange on the lattice.   
    \item For each consecutive edge $(e_{2i-1}, e_{2i})$ in $L$, sample a random number $p_e \sim U(0,1)$.   
    \item If $p_e\leq p$, swap the edges $e_{2i-1}=(u_{2i-1},v_{2i-1})$ and $e_{2i}=(u_{2i},v_{2i})$ to $e_{2i-1}^\prime=(u_{2i-1},v_{2i})$ and $e_{2i}^\prime=(u_{2i},v_{2i-1})$; otherwise, keep the edges as they are, $e_{2i-1}^\prime=e_{2i-1}$ and $e_{2i}^\prime=e_{2i}$.   
\end{itemize}
This yields a revised network, $G=(V,E^\prime)$, with $E^\prime \!=\! \{e_1^\prime, e_2^\prime,\dots,e_{|E|}^\prime\}$. Notably, these randomized network structures converge to the RRG as $p$ becomes ${\cal O}(1)$.

\section{Corrections to the steady-state  density} \label{appendix-ss}
Here we derive closed-form equations for $\alpha$ and $K$ in the main text, which allow one to determine the correction to the steady-state infection density for arbitrary $k_0$. 
Writing Eq.~\eqref{eq: F_ii ss} for shells $\ell=1,2$  yields 
\begin{eqnarray}
&& \bar{F}_{II}^{(1)} \!=\! \frac{\bar{F}_{II}^{(1)}\!-\!\bar{F}_{II}^{(1)}\bar{x}_{I}^{\text{MF}}}{\bar{F}_{II}^{(1)}-\bar{x}_{I}^{\text{MF}}}\!\! \left[p\left(1|1\right)\bar{F}_{II}^{(1)} \!+\! p\left(2|1\right)\bar{F}_{II}^{(2)}\right] \!+\! \frac{\bar{F}_{II}^{(1)}}{k_0 \bar{x}_{I}^{\text{MF}}}, \nonumber\\
    &&\bar{F}_{II}^{(2)} \!=\! \frac{\bar{F}_{II}^{(1)}\!-\!\bar{F}_{II}^{(2)}\bar{x}_{I}^{\text{MF}}}{\bar{F}_{II}^{(1)}-\bar{x}_{I}^{\text{MF}}} \!\!\left[p\!\left(1|2\right)\!\bar{F}_{II}^{(1)} \!+\!p\!\left(2|2\right)\!\bar{F}_{II}^{(2)} \!+\!p\!\left(3|2\right)\!\bar{F}_{II}^{(3)}\! \right]\!.\nonumber 
\end{eqnarray}
Substituting Eq.~\eqref{eq: F_ss} for shells $\ell=1,2,3$ into these equations, and using the fact that $1=p(k+1
|k) + p(k|k) + p(k-1|k)$, we obtain two coupled equations for $\alpha$ and $K$:
\begin{eqnarray} \label{eq:alpha_ss-appendix}
&&  \alpha K = -\frac{1-k_{0}\bar{x}_{I}^{\text{MF}}p\left(0|1\right)}{\frac{1-k_{0}\bar{x}_{I}^{\text{MF}}}{1-\bar{x}_{I}^{\text{MF}}}+k_{0}\bar{x}_{I}^{\text{MF}}\left(p\left(1|1\right)+p\left(2|1\right)K\right)},\\
&&0=p\left(3|2\right)\bar{x}_{I}^{\text{MF}}K^{3}+\left(\bar{x}_{I}^{\text{MF}}p\left(2|2\right)-p\left(3|2\right)\right)K^{2}\nonumber\\
&&\hphantom{=}+\left(p\left(1|2\right)\bar{x}_{I}^{\text{MF}}-p\left(2|2\right)+1\right)K-p\left(1|2\right) \nonumber \\
&&\hphantom{=}-\frac{1-\bar{x}_{I}^{\text{MF}}}{\alpha K}\left[p\left(3|2\right)K^{2}\!+\!p\left(2|2\right)K\!+\!p\left(1|2\right)\right]  + \frac{K\bar{x}_{I}^{\text{MF}}}{\alpha K}\!. \nonumber
\end{eqnarray}
Substituting 
$\alpha$ from the first of Eqs.~\eqref{eq:alpha_ss-appendix}, yields a cubic equation for $K$, $AK^3\!+\!BK^2\!+\!CK\!+\!D\!=\!0$, with 
\begin{align}
    A &=k_{0}(\bar{x}_{I}^{\text{MF}})^2p\left(3|2\right)\left(1 - p\left(1|1\right)\right) \nonumber \\
    &\hphantom{=} -k_{0}\bar{x}_{I}^{\text{MF}}p\left(3|2\right)p\left(2|1\right) -p\left(3|2\right), \nonumber\\
    B &= k_{0}(\bar{x}_{I}^{\text{MF}})^2\left[p\left(2|2\right)-p\left(1|1\right)p\left(2|2\right)+p\left(1|1\right)p\left(3|2\right)\right] \nonumber \\
    &\hphantom{=}+k_{0}\bar{x}_{I}^{\text{MF}}p\left(2|1\right)\left[1+p\left(3|2\right)-p\left(2|2\right)\right] -p\left(2|2\right)\bar{x}_{I}^{\text{MF}}, \nonumber\\
    C &=k_{0}(\bar{x}_{I}^{\text{MF}})^2\left[p\left(1|2\right)-p\left(1|1\right)p\left(1|2\right)+p\left(1|1\right)p\left(2|2\right)\right] \nonumber \\
    &\hphantom{=}+k_{0}\bar{x}_{I}^{\text{MF}}\left[1-p\left(2|1\right)\left(1-p\left(2|2\right)+p\left(1|2\right)\right)\right] \nonumber \\
    &\hphantom{=}-p\left(1|2\right)\bar{x}_{I}^{\text{MF}}+\bar{x}_{I}^{\text{MF}}\left(1-k_{0}\right)/(1-\bar{x}_{I}^{\text{MF}}), \nonumber\\
    D &= k_{0}(\bar{x}_{I}^{\text{MF}})^2p\left(1|1\right)p\left(1|2\right)+k_{0}\bar{x}_{I}^{\text{MF}}p\left(2|1\right)p\left(1|2\right).\nonumber \label{eq: steady-state end-appendix}
\end{align}

\textcolor{black}{\section{Comparison with the SPM} \label{appendix-c}}

\textcolor{black}{The standard pairwise model (SPM)~\cite{keeling_effects_1999, kiss2017mathematics} is a well-established and widely accepted closure scheme in network epidemiology, offering a major improvement over mean-field approaches by explicitly accounting for local network structure through nearest-neighbor correlations. 
The SPM yields the following steady-state value: 
$\bar{x}_I^{\text{SPM}} \!=\! 1 \!-\! 1/\left[\Lambda \!-\! 1/(k_0 \!-\! 1)\right]$, where $\Lambda=\beta k_0/\gamma$. Expansion of $\bar{x}_I^{\text{SPM}}$ in powers of  $k_0^{-1}\!\ll\! 1$ yields
\begin{equation}
\bar{x}_I^{\text{SPM}} \simeq \bar{x}_I^{\text{MF}} - \frac{1}{k_0\Lambda^2}-\frac{\Lambda^2+\Lambda}{k_0^2\Lambda^4},
\end{equation}
up to $\mathcal{O}(k_0^{-3})$ corrections.
The $\mathcal{O}(k_0^{-1})$ term agrees with our MPM result [Eq.~\eqref{x-bar-power-series}], yet the ${\cal O}(k_0^{-2})$ term differs. This reflects finite-degree clustering effects that are not captured by the SPM and is precisely what leads to the smaller errors of the MPM steady-state predictions.}

\begin{figure}[h!]
\centering
\includegraphics[width=\linewidth]{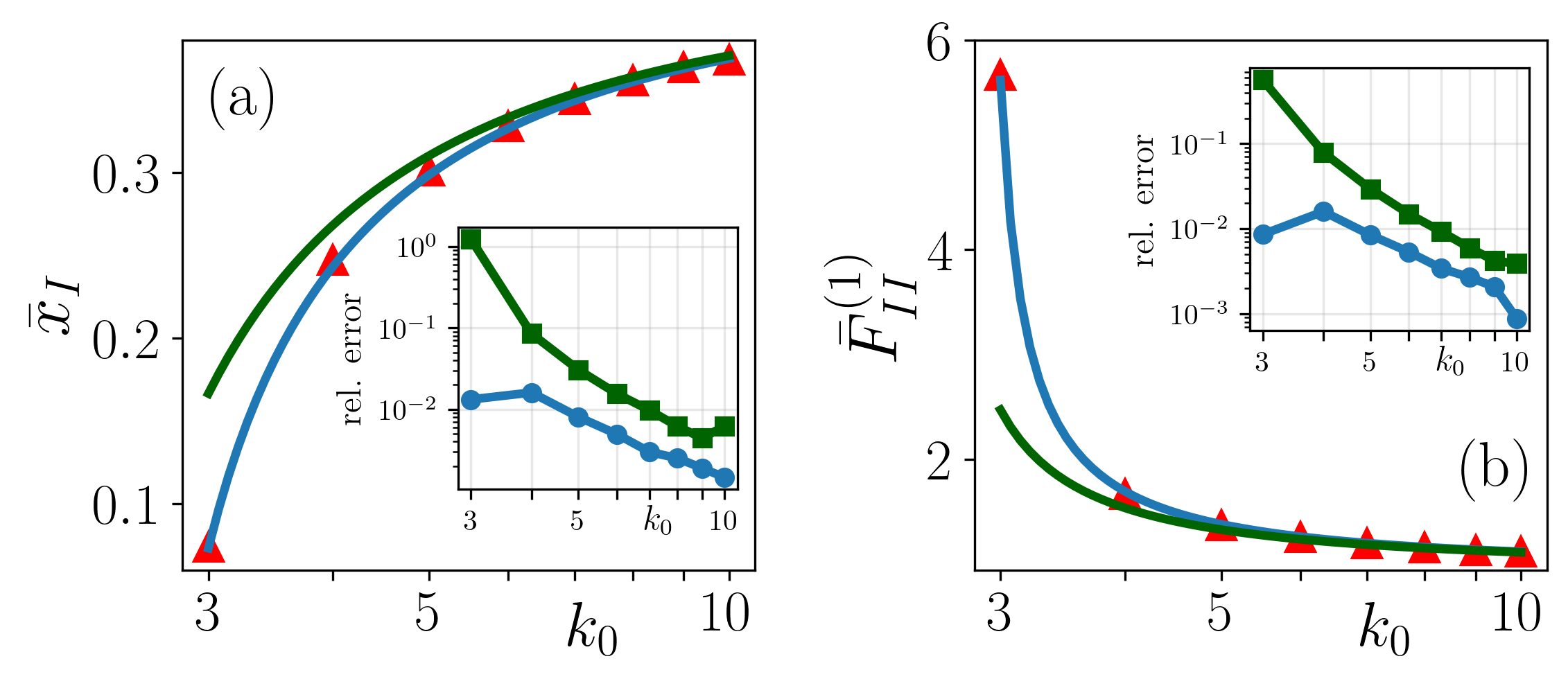}
\vspace{-8mm}\caption{(a) Steady-state infection density $\bar{x}_I$ and (b) correlation function $\bar{F}_{II}^{(1)}$ versus $k_0$ for $\Lambda=1.7$. 
Shown are the MPM (blue line), SPM (dark green line) and numerical simulations (red triangles) with $N=10^4$ nodes. Insets show relative error between simulations and MPM (circles), and SPM (squares).}
\label{fig:appendix_b_comparison}
\end{figure}

\textcolor{black}{Figure~\ref{fig:appendix_b_comparison}(a) presents the steady-state prevalence $\bar{x}_I$ across a range of mean degrees $k_0$ for $\Lambda=1.7$. The MPM with $\ell_{\max}>1$ (blue) agrees well with simulation data (red triangles), with deviations around $1\!-\!2\%$ over the entire range of $k_0$. In contrast, the SPM (dark green line) systematically overestimates the prevalence, with large errors for $k_0={\cal O}(1)$. This reflects the contribution of higher-order spatial correlations captured by evaluating the closure at multiple shell distances.
Figure~\ref{fig:appendix_b_comparison}(b) shows the corresponding steady-state  distance-one pairwise correlation function $\bar{F}_{II}^{(1)}$. Here, the MPM again agrees well with simulations, while the SPM greatly underestimates $\bar{F}_{II}^{(1)}$ for $k_0={\cal O}(1)$. }

\end{document}